\documentclass[11pt]{article}
\usepackage{amsmath}
\usepackage{amsthm,amssymb}
\usepackage{color}
\usepackage{authblk}
\usepackage[numbers]{natbib}
\usepackage[T1]{fontenc}
\usepackage[utf8]{inputenc} 
\usepackage{subfig}
\usepackage{graphicx}

\usepackage{listings}
\date{}

\title{\Large \bf  Hybrid Galam--Bass Model for Technology Innovation}
\author[1]{Giulia Rotundo }
\author[,2,3]{Roy Cerqueti \thanks{Corresponding author email: roy.cerqueti@uniroma1.it}}
\author[4,5]{Gurjeet Dhesi }
\author[4,5]{Claudiu Herteliu }
\author[6,5]{Parmjit Kaur }
\author[7,8,5,6]{Marcel Ausloos}

 \bigskip  
 
 \affil[1] { Department of Statistical Sciences, Sapienza University of Rome, p.le A. Moro 5, 00185 Rome, Italy;  email: giulia.rotundo@uniroma1.it}
 
 \affil[2]{ Department of Economics and Social Sciences, Sapienza University of Rome, p.le A. Moro 5, 00185 Rome, Italy;  email:  roy.cerqueti@uniroma1.it }
 
\affil[3]{ Groupe de Recherche  Angevin en Économie et Management (GRANEM), University of Angers, SFR Confluences,  F-49000 Angers, {France}}

\affil[4]{ Department of Statistics and Econometrics, Bucharest University of Economic Studies, Calea Dorobantilor 15-17, 010552 Sector 1, Bucharest, Romania;  email:  dhesig74@gmail.com (G.D.); hertz@csie.ase.ro (C.H.) }

\affil[5]{ Department of Statistics, Predictions and Mathematics, Universitatea Babe{\c{s}}-Bolyai, Str. Mihail Kogălniceanu 1, 400084, Cluj-Napoca, Romania   }

\affil[6]{ Guildhall School of Business and Law, London Metropolitan University, London N7 8DB, UK;    email : p.kaur@londonmet.ac.uk  }
  
\affil[7]{ School of Business, University of  Leicester, Brookfield, Leicester,   LE2 1RQ, UK;   email : ma683@leicester.ac.uk}  

\affil[8]{ Group of Researchers for Applications of Physics in Economy and Sociology (GRAPES), Rue de la belle jardini\`ere, 483,  B-4031 Angleur, Li\`ege, Belgium; email: marcel.ausloos@uliege.be} 
\begin{document}
\maketitle  

\begin{abstract}  
This work proposes a hybrid model that combines the Galam model of opinion dynamics with the Bass diffusion model used in technology adoption on Barabasi--Albert complex networks. The main idea is to advance a version of the Bass model that can suitably describe an opinion formation context while introducing irreversible transitions from group \( B \) (opponents) to group \( A \) (supporters). Moreover, we extend the model to take into account the presence of a charismatic competitor, which fosters conversion back to the old technology. The approach is different from the introduction of a mean field due to the interactions driven by the network structure. Additionally, we introduce the Kolmogorov--Sinai entropy to quantify the system's unpredictability and information loss over time. The results show an increase in the regularity of the trajectories as the preferential attachment parameter increases.
\end{abstract}


\section{Introduction}
	
	The statement  {\it {technological innovation cannot be halted}
} is often attributed to Bill Gates, the co-founder of Microsoft. He is well-known for his strong thoughts on technology's progress and its potential to dramatically transform industries and impact global resources.
	Is this really true? Yes and no. In Section \ref{sec:mot}, we present some instances where such a sentence is discussed. 
	
	In general, it is interesting to understand how a minority can shift to a majority. Evidently, some factors intervene in this topic. 
	This paper enters this debate.
	
	We build on 
	 the work of the prominent physicist Serge Galam on sociophysics,
	who is a scientist applying the principles of physics to social phenomena. In his work, Galam contributes to exploring collective human behaviors, mainly in contexts like opinion dynamics, the development of social networks, and political flows~\cite{Galam2002,Galam2011}.
		
	A major contribution 
	 of Galam  modeling the spread of opinion---including voting models---is the explanation of the emergence of consensus or polarization due to the meeting of small groups of individuals, where the outcome of each single group is driven by the majority rule~\cite{Galam2002,Galam1986}. The opinion of each individual may change, due to local interactions or to external influences. An aspect that is widely examined is the influence of a small number of contrarians, whose opinion is just the opposite of that of the majority~\cite{Galam1986,Galam2004}. When does an initial minority take over the majority? 
	
	The models are quite powerful and well-suited for cases of studying the spread of innovations~\cite{Galam2011}, considered as an implementation of the shift of opinions, and they can be extended to practical cases of everyday life---for instance, to the diffusion of goods and behaviors~\cite{Galam2005}.
	An example reported by Galam during his talks at a number of workshops is immediately evident to flight passengers: due to some severe episodes and outcomes related to peanut allergy, which pertains to a minority of the world population, flight companies are more keen to offer crackers as snacks rather than peanuts; so, a small group is driving the choice of snacks for all the population. Within this context, the adoption of crackers takes the place of the new technology, as opposed to peanuts, which represent the old technology.
	The interactions and the grouping can be either at random or driven by a (complex) network structure. 
	Network theory and statistical mechanics, in fact, are used  to  set up and analyze such social phenomena, leading to the term sociophysics to represent a true bridge between the natural and social sciences~\cite{Galam2002,Galam2011,Galam1986,Galam2005}.
	
	%%%%%%%%%%%%%%%%%%
	In this work, we contribute to the study of Galam's model applications by adding the classic Bass model for the diffusion of technologies, which describes how a new product or innovation spreads through a population over time \cite{Bass2004,Mahajan1990,tigert,VRAC2015}.
	%Practically, in terms of the Galam model.
	The Bass model focuses on two groups of individuals: innovators (early adopters) and imitators (those who adopt based on the influence of others). Innovators adopt the technology independently, while imitators are influenced by the number of people who have already adopted the innovation. The model suggests that the adoption rate depends on two main parameters: the coefficient of innovation (how likely innovators are to adopt) and the coefficient of imitation (how much influence the adopters have on others).
	
	The Bass model captures the S-shaped curve typically seen in the diffusion process, with slow initial adoption, a rapid growth phase, and, eventually, a slowing down as the market becomes saturated.
	
	A key difference between the models is that in the Bass model there is no step back to the old technology. The spread of the new technology can only increase until it reaches $100\%$.
   
	All the models are run on Barabasi--Albert (BA) networks. 
	BA networks are scale-free networks generated using the preferential attachment mechanism, where new nodes are added to the network with a probability proportional to the number of connections (degree) of the existing nodes. This results in a small number of highly connected nodes (hubs) and a large number of low-degree nodes, characteristic of real-world complex networks, such as social networks, the internet, and biological networks.
	
	In the context of social systems, BA networks are particularly useful because they capture the emergence of hubs (influential individuals or entities) and the power law distribution of node degrees often observed in networks drawn from empirical datasets. The BA model provides insights into the dynamics of influence spread, information dissemination, and social interactions. It is widely used to model how innovations or behaviors propagate within a population, especially in scenarios where individuals are more likely to adopt new behaviors from their already connected peers~\cite{barabasi1999-1,Vespignani2006-1,Barabasi2002}.
Since the diffusion driven by the Bass model always reaches $100\%$ of the network nodes, it is useless to focus on  the final number of involved units  and the time span, as in other diffusion models on networks. Instead, we keep track of the randomness of the trajectories. The tool selected for this purpose is the Kolmogorov--Sinai entropy~\cite{kolmogorov1958invariant,sinai1959entropy}. Of course, other measures of regularity could have been selected, such as the the Hausdorff dimension or Hurst's $H$ index, but they are not suitable in our context. Indeed, the former finds its best usage for characterizing the geometric complexity in the phase space of attractors, and the latter is specific for static long-term correlations rather than for evolution over time.
	
	%We are going to elaborate on the models, and on the thresholds which lead to different dynamics. 
%	The Kolmogorov-Sinai entropy is used to get information on the randomness of the trajectories. 
    %Other models which do not necessarily lead to $100\%$ of involved units,
    
	The remaining sections are organized as follows: Section \ref{sec:mot} outlines the main features of the models used; Section \ref{sec3} shows the results; our conclusions  follow. %EE: Attention: Please note that this summary does not accurately describe the following sections in the paper.
	
	\section{Motivating Examples}
	\label{sec:mot}
	Let us consider three examples. 
	
	The first one is about the adoption and evolution of USB standards. Such innovation has played a key role in the storage of data on external devices. On this, see~\cite{giersch2016usbc-1} and ~\cite{benedict2009usb-1}, where the latter paper discusses the exploitation of USB Type-C. 
	Interestingly,   
\cite{buchanan2018role-1} emphasizes the role that USB plays in contributing to unified connectivity, and   \cite{sobel2017usbc-1} elaborates on the future of digital plug standards, with a focus on USB-C. 
	Additionally,  ~\cite{berkman2019usb-1} examines the effect of the adoption of the standard USB, with a special mention of the storage of personal health records, while ~\cite{barrett2020usb-1} discusses the shift from USB 1.0 to USB-C in the context of safety.
	It is worth noting that the adoption of this new technology does not lead to a possible step back to the previously adopted ones.
	
	Let us consider a different example. Electric vehicles---highly popular nowadays---were used well before the current period. At the beginning of the XX century, a large part of the vehicles in the United States were electric (see, e.g.,~\cite{sivak2014early-1,hirsch1999rise-1}). Electric vehicles were replaced by gasoline cars, in light of the development of the fossil fuels market %EE: Please check intended meaning has been retained.
	 (see, e.g.,~\cite{buchanan2009electric-1}).  This is a paradigmatic example of a minority usage of technology becoming a majority. In recent times, we have returned to the old technology of electric cars, which is increasingly popular in Western countries. 
	
	The third example relates to the adoption of green technologies. 
	It is known that the United States has withdrawn from the agreements on climate change, which are focused on promoting green energy. This decision was made by President Trump on January 20th, 2025. It has raised criticisms and concerns, mainly in the context of the Paris Agreement %and Just Energy Transition Partnership (JETP),
	but also in the context of the entire socio-economic reality. %the Just Energy Transition Partnership (JETP), and funding for climate finance.
	% In details, on January 20, 2025, President Trump signed an executive order that directed the immediate withdrawal of the United States from the Paris Agreement. 
	Indeed, this decision represents a remarkable step back in the fight against the detrimental effects of climate change and might have effects also on US economic competitiveness~\cite{morganlewis2025-1}.
	Furthermore, in March 2025 the United States also withdrew from the Just Energy Transition Partnership (JETP),  which was an institutional instrument to support developing countries in shifting from fossil fuels to renewable  energy; see  https://climate.ec.europa.eu/news-your-voice/news/joint-statement-international-partners-group-us-withdrawal-just-energy-transition-partnership-south-2025-03-19$_{-}$en.
	
	%\footnote{\url{https://www.uci.org/regulations/3MyLDDrwJCJJ0BGGOFzOat}}
	
	In general, President Trump has implemented a strong reduction in the budget for environment-targeted actions in light of climate change mitigation~\cite{guardian2025-1}.
	%These withdrawals have been met with widespread criticism from international leaders and environmental groups, raising concerns about the U.S. retreating from its leadership role in global climate efforts. However, these actions are shaping the future of international climate policy and funding.
	
	The question of interest is: Will the actions of the President of the United States lead to a change in the behavior of other countries in the context of environmental protection and the fight against climate change?
	This is a clear example of the shift to an old technology due to the presence of a ``charismatic competitor'', who can individually reverse the decisions and preferences of other agents. 
	Indeed, the US is an undoubtedly relevant country playing a leading role even when it starts from a minority position.

	\section{Methodological Basis of the Model} \label{sec3}
	
	The presented model is based on four main pillars. We specifically refer to the Bass model~\cite{Bass2004,Bass1969}, the Galam model~\cite{Galam2020}, and the Kolmogorov--Sinai entropy~\cite{kolmogorov1958invariant,sinai1959entropy}, which is of great usefulness for detecting the regularity of trajectories. Furthermore, we assume that individuals have social connections. Thus, individuals are nodes of a network with links describing the social interconnections between them. For this reason, we need complex networks and we refer specifically to the Barab\'{a}si--Albert (BA) networks~\cite{barabasi1999-1}. %This section provides the main ingredients and some information on these concepts.
		
	\subsection{Bass Diffusion Model}
	The Bass model is one of the most popular instruments for describing the adoption of a new technology. Its theoretical conceptualization moves from a Cauchy problem, as~follows,	
	\begin{equation}
		\label{bass}
		\begin{cases}
			\frac{dF(t)}{dt} = p (1 - F(t)) + q F(t) (1 - F(t)), \qquad t >0
			\\
			\mbox{} \\
			F(0)=0
		\end{cases} 
	\end{equation}
	where $t \geq 0$ represents time, \( F(t) \) measures the cumulative fraction of individuals who adopt the considered technology, and, as a consequence, \( (1 - F(t)) \) represents the fraction of individuals not adopting the new technology at time \( t \). The parameter \( p \) describes the innovators---who are independent adopters---while \( q \) is a parameter associated with the imitators, who adopt the new technology on the ground of social influence. 
	
Stating $F(0)=0$ means that at the initial time $t=0$ no individuals adopt the new technology, as the evidence clearly suggests. 

	\subsection{Galam Model}
The Galam model moves from two groups of individuals, namely, group $A$ and group~$B$. The elements of group $A$ are supporters of the new product, while the individuals in group~$B$ are opponents and may switch to group $A$ on the basis of their social interactions.
	
%	As already said above, agents are the nodes of a network.
The role played by the interactions depends on the specific agents influenced %EE: Please check intended meaning has been retained.
 by such interconnections. On this basis, we can identify different groups of individuals:
%	The presence of interactions allows to define some behaviors of the individuals. Hence, we have special groups of agents.
	%A further characteristic of each agent (in our case, node) of the network is the possibility to be a floater
	
	\begin{itemize}
		\item {\it {Floaters}
}: floaters are individuals belonging to group $B$ who can change their stances as a consequence of interactions.
	\end{itemize}
  
	Importantly, in the first step in our study, we assume that going back from the adoption of the new technology is not possible. Thus, floaters belong to $B$ and can move to $A$, and there are no people moving from $A$ to $B$. 
	
	%At a first step, when there are no withdrawal from the adoption of the new technology, we consider that the floaters can only move from B to A, not viceversa.
	\begin{itemize}
		\item {\it {Stubborn agents}
}: 
        stubborn agents are individuals in a troupe who cannot change opinion and shift to the other group.
%        stubborn agents are individuals in group $A$ who cannot change and shift to $B$.
	\end{itemize}
	
	In the original setting proposed by~\cite{Galam2020}, one can find also the contrarians:
	\begin{itemize}
		\item {\it {Contrarian agents}
}: the contrarians are floaters who decide to act in opposition %EE: Please check intended meaning has been retained.
 to the majority group they contribute to, by changing opinion.% once groups are dismantled.%, decide individually to shift opinion to oppose the group majority they contributed to.
	\end{itemize}
	
	As we will see, in the first step of our study we also assume that there are no contrarian~agents. %for the sake of simplicity. % we present assumes that contrarians and floaters coincide
	
	%In our first setting, we unify the property "contrarians" to the one of "floaters" since the floating can only happen from A to B.

   The wide work of Serge Galam proposes many variants and details of the evolution of the proportion of elements of type $A$ versus elements of type $B$. 
 One characteristic is to consider changes of opinion due to internal comparison within small  groups.  
  For instance, in~\cite{galam2023unanimity} a case of groups composed of four elements is detailed in Equation (\ref{bass}): $p_1$ represents the proportion of agents holding opinion $A$, and its value after one cycle is \mbox{given by}
\begin{equation}
p_1 = p_0^4 + 4 p_0^3 (1 - p_0) + 3 p_0^2 (1 - p_0)^2.
\end{equation}
where $p_0$ is the initial proportion of agents holding opinion $A$ before the update, 
 $p_1$ is the new proportion of agents holding opinion $A$ after one update, and the three terms have the following meaning: $p_0^4$ is the probability that all 4 agents in a discussion group initially support $A$ (thus, it remains $A$); $4 p_0^3 (1 - p_0)$ is the probability that 3 support $A$ and 1 supports $B$ (majority is $A$);  $3 p_0^2 (1 - p_0)^2$ is the probability of a tie (2 support $A$ and 2 support $B$); under the assumption here, ties do not change opinions and stay as they are.

The above is just one out of the many equations that describe the evolution of the number of elements in group A (and, consequently, in B), since the calculus is repeated over groups of different sizes.
For instance, an evolution equation reported in~\cite{jacobs2019two} is 
\[
p_{t+1} 
= f_1(\alpha_A, \alpha_B, \gamma_A, \gamma_B; p_t) 
= \alpha_A + (1 - \gamma_A)(p_t - \alpha_A) + \gamma_B (1 - \alpha_B - p_t)
\]

\noindent
where 
$p_t$ is the proportion of agents holding opinion $A$ at time $t$;
$\alpha_A$ is the fraction of stubborn (inflexible) $A$ agents who never change opinion;
$\alpha_B$ is the fraction of stubborn (inflexible) $B$ agents who never change opinion;
$\gamma_A$ is the fraction of contrarian $A$ floaters who switch from $A$ to $B$;
$\gamma_B$ is the fraction of contrarian $B$ floaters who switch from $B$ to $A$.

	Furthermore, as already said above, agents change their opinion on the basis of their adjacents in the network. In~\cite{li2023graph}, the experiment is set up on networks, although no recurrent formulas are highlighted. 
Our model is based on ideas and concepts proposed through the scientific production of Serge Galam and we propose a view at the macro level without entering into the details of the single equations.
        
    We refer to Section \ref{sec:res} for the details of the proposed numerical validation of the theoretical model.
			
	\subsection{Kolmogorov--Sinai Entropy}
	We compute the Kolmogorov--Sinai (K--S) entropy to provide a measure of system unpredictability. Indeed, K--S quantifies the time-dependent information loss.
	
	For a given partition \( \mathcal{P}=\{P_A,P_B\} \) of the set of the agents---where $P_A$ and $P_B$ account for the agents in the group, either  $A$ or $B$, respectively---we compute the entropy of $\mathcal{P}$ as follows:
	
	\begin{equation}
		H_{\mu}(\mathcal{P}) = -\left[ \mu(P_A) \log \mu(P_A)+ \mu(P_B) \log \mu(P_B) \right]
	\end{equation}	
	 where  
 \( \mu(P_A) \) and $\mu(P_B)$ are the probability that the agents are of type \( A \) and $B$, respectively---or, equivalently, they belong to group $A$ and $B$, respectively. 
	Such a partition generates a dynamical system where a given agent $x$ that is in $P_A$ at time $i$ can be viewed as $x \in f^{-i}(P_A)$, and the same for $P_B$.  
	 The K--S entropy is
	
	\begin{equation}
		h_{\mu}(f) = \sup_{\mathcal{P}} \lim_{n \to \infty} \frac{1}{n} H_{\mu} \left( \bigvee_{i=0}^{n-1} f^{-i} \mathcal{P} \right).
	\end{equation}
	
	%For our discrete system, we approximate entropy using the probability distribution of agents in \( A \) and \( B \) at each time step.

The K--S entropy is the joint Shannon entropy of the random variables that describe the system as a function of time. Thus, the K--S formula sets an explicit measure of the temporal regularity or randomness of the dynamics under study. Actually, it calculates the amount of new information produced at each time step. The less new information, the more predictable the system and the less uncertainty in the future.  
In inhomogeneous networks, such as the BA ones, the regularity of the trajectories is all-but-obvious, due to the strong dependence, on top of all the other parameters, on the sequence of high versus low connected nodes, which is met in any simulation run, and their state. 

	\subsection{The BA Networks}
	The Barabási--Albert (BA) model generates networks of the scale-free type by implementing a preferential attachment procedure. New nodes are more likely to connect to other existing nodes having higher degrees. According to this mechanism, one has the formation of hubs---i.e., nodes having an extraordinarily large number of connections---while the greater part of the nodes have fewer connections. As an outcome, one has a power law distribution for the nodes degree. A preferential attachment parameter \( m \) drives the probability that a node has a specific degree, and this affects the rate at which new connections are created. Thus, higher values of \( m \) are associated with faster growth of the hubs and a more evident skewness in the distribution of the degree. The BA networks are often associated with the modeling of real-world phenomena such as the distribution of influence in social networks, which is evidently uneven. Such a network model is often associated with network dynamics and the spread of information in complex systems. For more details, please refer to~\cite{barabasi1999-1,Barabasi2002,pastor2002immunization}.

	\section{Model Definitions, Methods, and Software Implementation}
	
	\subsection{Hybrid Galam--Bass Model with No Back-Switch to the Old Technology}
	By mapping the Bass parameters to Galam's framework, we define the following:
	
	\begin{itemize}
		\item \( F(t) \equiv A(t) \) (the fraction of group $A$ agents who are stubborn).
        		\item \( 1 - F(t) \equiv B(t) \) (the fraction of agents still in $B$ who are floaters).
		\item \( p \equiv k \) (spontaneous conversion from B to A).
		%\item \( q \) corresponds to floaters in group B who switch based on majority influence.
        \item \( q \) is the imitation coefficient, which modulates the strength of the social influence of group $A$ over group $B$, and represents the influence of floaters in B transitioning based on social interactions.
	\end{itemize}

    % It is worth to remark that the parameter $q$ modulates the strength of the social influence of the group $A$ over the group $B$. The higher are the values of $q$, the easier is the conversion of the floaters is $B$ to $A$.

	The hybrid model is then formulated as
	
	\begin{equation} \label{hybrid}
		\frac{dA(t)}{dt} = k B(t) + q A(t) B(t).
	\end{equation}
	
The key assumption is that ``once an agent transitions to A, they cannot switch back to~B''. 

    In discrete time, Equation (\ref{hybrid}) reads as 

    	\begin{equation}  
		A(t+1)-A(t) = k B(t) + q A(t) B(t)
	\end{equation}
	
	Note that when the agents are connected through a network, the interaction is only through the neighbors; that is, the term $A(t)B(t)$ is considered only for nodes directly connected in the network due to the presence of a network link. In this, the formulation differs from the one in paper~\cite{Galam2020}.
Therefore, the process of imitation of %EE: Please check intended meaning has been retained.
the global adoption level $A(t)$  is only a rough approximation of the local fraction of adopting neighbors, which holds in the case that the network is complete.
    In detail, the 
    evolution of the fraction of adopters $A(t)$ in a network-based discrete-time Bass model can be written as
\begin{equation}
 \label{BassDiscreto}
A(t+1) - A(t)
= kB(t) + q   \bar{A}_{\text{neighbors}}(t) B(t)
\end{equation}
where $\bar{A}_{\text{neighbors}}(t)$ is the fraction of agents in group $A$ who are adjacent---in terms of the considered network---to an agent in group $B$.
%\[
%\bar{A}_{\text{neighbors}}(t)= \frac{1}{N_B} \sum_{i: S_i(t)=0} \frac{\sum_{j \in \mathcal{N}(i)} S_j(t)}{|\mathcal{N}(i)|}
%\]
%represents the average fraction of adopted neighbors among all non-adopters, with
%\begin{itemize}
%    \item $S_i(t) = 1$ if node $i$ is in the set of adopters at time $t$, and $0$ otherwise,
%    \item $\mathcal{N}(i)$ the set of neighbors of node $i$,
%    \item $N_B$ the number of non-adopters at time $t$.
%\end{itemize}

\noindent
Since the transition is only from B to A,   the presence of a percentage $\alpha$ of contrarians is included in the spontaneous conversion from B to A and taken into account in $\hat{k}=(k+\alpha)$. Equation (\ref{BassDiscreto}) becomes
\begin{equation}
A(t+1) - A(t)
= \hat{k} B(t) + q   \bar{A}_{\text{neighbors}}(t) B(t)
\end{equation}
\noindent
We examine the temporal evolution of $A(t)$ via numerical simulations. 
	Algorithm \ref{algo1} outlines the steps for the simulation of the hybrid Galam--Bass model on BA networks.
	Each BA network contains 1000 nodes, and it is %EE: Please check intended meaning has been retained.
	 generated with a preferential attachment parameter $m \in m_{\text{values}}= \{1, 3, 5, 10, 20\}$ starting from a seed of 
	$m_0 = 5$ nodes. Moreover, in line with~\cite{bertotti2019bass}, the value $q = 0.3$ of the influence-driven adoption (imitators) corresponds to a 
     realistic value in innovation diffusion theory~\cite{jiang2006virtual}, and    
    $k = 0.02$ is the probability of independent adoption (innovators).

    We set
	$\alpha = 0.1$ as the fraction of the contrarians (i.e., randomly selected so as to be $0.1\%$ of the total number of agents),
	$T_{max} = 50$ is the number of simulations measured in terms of time steps,
	$numNetworks = 100$ is the  number of  networks generated for each value of $m$, and
	$numSimulations = 100$ is the number of simulations per network.
 
 \bigskip

=========================================

%\subsection
{\bf Algorithm 1. }Bass model on Barabási--Albert networks with contrarians and Kolmogorov--Sinai entropy. \label{algo1}

%\begin{algorithm}
--------------------------------------------------------------------------------------------

%\begin{algorithmic}
    %\State 
    \textbf{Input:} Parameters, including the number of nodes $N$, the initial BA parameters $m_0$, the adoption probabilities $p$ and $q$, the contrarian fraction $\alpha$, the number of time steps $T_{\max}$, the array of $m_{\text{values}}$, the number of networks $numNetworks$ for each value of $m$, and the number of simulations $numSimulations$ (run of the model) on each network.
  
\textbf{Output:} Adoption rates over time and Kolmogorov--Sinai entropies (for different network configurations).
    
\textbf{Begin}: 

For each value of $m$ in $m_{\text{values}}$ do the following:
     	
	 For each network (from $1$ to $numNetworks$) do the following:
 
  \;\;\;Generate a Barabási--Albert network with parameters $N$, $m_0$, and $m$.
  
  \;\;\;Compute the adjacency matrix of the network.
    
    \;\;\;For each simulation (from $1$ to $numSimulations$) do the following:
    
   \;\;\;\;Initialize all nodes as non-adopters.
  
     \;\;\;\;Randomly designate a fraction $\alpha$ of nodes as contrarians.
   
     \;\;\;\;Seed initial innovators who adopt independently with probability $p$.
    
   \;\;\;\;For each time step $t = 1$ to $T_{\max}$ do the following:
       
  \;\;\;\;\;\;\;Create a copy of the current adoption states for synchronous updates.

 \;\;\;\;\;\;\;For each node $i$ that is still a non-adopter do the following:
 
 \;\;\;\;\;\;\;\;\;Find the neighbors. %Identify its neighbors and compute the fraction who have adopted.
  
  \;\;\;\;\;\;\;\;\;Set probability based on contrarian behavior.%If node $i$ is a contrarian then:
 
  \;\;\;\;\;\;\;\; \;\;\;\;Set probability based on influence adopted. %Set adoption probability to $p \times (1 - f_i)$ to reduce influence from adopters.
   % \State \hspace{5em} Else:
    %\State \hspace{6em} Set adoption probability to $p + q \times f_i$ to reflect imitation behavior.
    
  %  \State \hspace{5em}  \;\;\;\;\;\;\;\;\; 
     \;\;\;\;\;\;\;\;\;If node $i$ adopts based on the probabilities above do the following:
    
  \;\;\;\;\;\;\;\; \;\;\;\;Update $i$'s state to adopter.
  
  %  \State \hspace{4em}   \;\;\;\;\;\;\;
      \;\;\;\;\;\;\;Record the current fraction of adopters.
    
 %   \State \hspace{3em}   \;\;\;\;\;\;\; 
         \;\;\;\;Compute the Kolmogorov--Sinai entropy of the adoption trajectory.
         
         \;\;\;\;Store the adoption rates and entropy for this simulation.

      Compute averages and standard deviations of the adoption rates and entropies across all runs.
   % \State Plot adoption rates versus time and entropy versus $m$.
  
    \textbf{End}
    
%\end{algorithmic}
%\end{algorithm}
  
  ========================================
  
	\subsection{Hybrid Galam--Bass Model with a Charismatic Competitor}
	
	The concept of the ``charismatic competitor'' has been introduced within the context of electoral dynamics~\cite{javarone2013network}. This competitor type leverages personal charisma to influence voter behavior and gain an electoral advantage. The research models election campaigns as complex networks, emphasizing how a charismatic individual's appeal can significantly alter the structure and flow of information within these networks. By attracting and mobilizing supporters through personal magnetism, the charismatic competitor can create a ripple effect, enhancing their visibility and influence across the electorate. The study highlights that such competitors often rely less on traditional campaign strategies and more on direct engagement and emotional connection with voters. This approach can lead to rapid shifts in public opinion, demonstrating the potent role that charisma plays in political competitions.
	Yet, although there are some conceptual similarities between the charismatic competitor and the mean-field effect, they operate in different ways: the charismatic competitor represents a real-world agent with direct interactions, while the mean-field effect is an abstract statistical assumption that smooths out individual variations. This difference can be emphasized by adopting specific views: 
	\begin{itemize}
		\item 
		Influence on the system. 
		The charismatic competitor in election campaigns shapes voter behavior by leveraging personal appeal and direct engagement, influencing the network dynamically. 
		The mean-field effect in physics and opinion dynamics represents how each individual is influenced by the average state of the whole system, smoothing out local fluctuations.
		\item Role in opinion formation. 
		A charismatic competitor acts as a focal point, influencing many individuals and triggering cascades of opinion changes.
		The mean-field effect assumes that individuals are exposed to an averaged global opinion rather than localized interactions.
		\item
		Global vs. local influence.
		The charismatic competitor creates a localized but strong effect that spreads outward, potentially reshaping the whole system.
		The mean-field effect assumes a uniform influence across the entire population without distinct \mbox{focal points.}
	\end{itemize}

	\section{Results and Interpretation}
	\label{sec:res}

	The simulation tracks the fraction of individuals in group A (supporters) and group B (opponents) over time. Due to the irreversible transition rule, the system always evolves toward full adoption (\( A \to 1 \)), with the speed of convergence depending on \( p \) and \( q \).
	
	Additionally, the Kolmogorov--Sinai entropy provides insight into the uncertainty of the system. At the beginning, entropy is high, due to a more balanced distribution of opinions. As the system converges towards full adoption, entropy decreases, reflecting reduced uncertainty.

	\subsection{First Case: No Back-Switch to the Old Technology}
\begin{figure}[]
		%il programma per generare queste figure si chiama BassMeglio100BA10simKS.m
	%	\centering
	\caption{ Hybrid  
 Bass--Galam diffusion model on BA networks with preferential attachment parameters $m=1,3,5,  10, 20$. Left side: the diffusion as a function of time. Each continuous curve is averaged over $100$ networks and $100$ simulations for each network. The dashed curves represent the intervals around the continuous curve due to the standard deviation. Right side: the Kolmogorov--Sinai. After the peak at $m=3$, the higher $m$ is, the faster the diffusion, and the trajectory is more regular.}
		\label{fig:Hybrid}
		\includegraphics[width=0.60\linewidth]{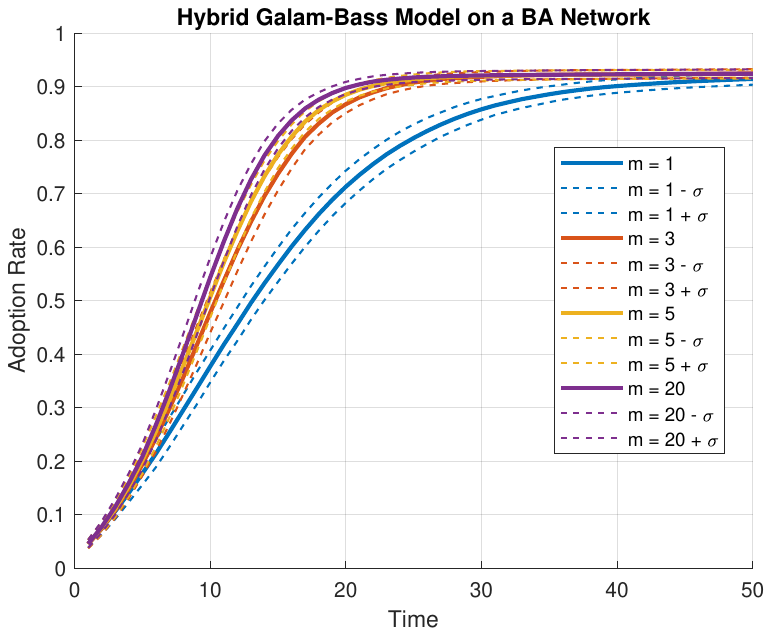} \includegraphics[width=0.60\linewidth]{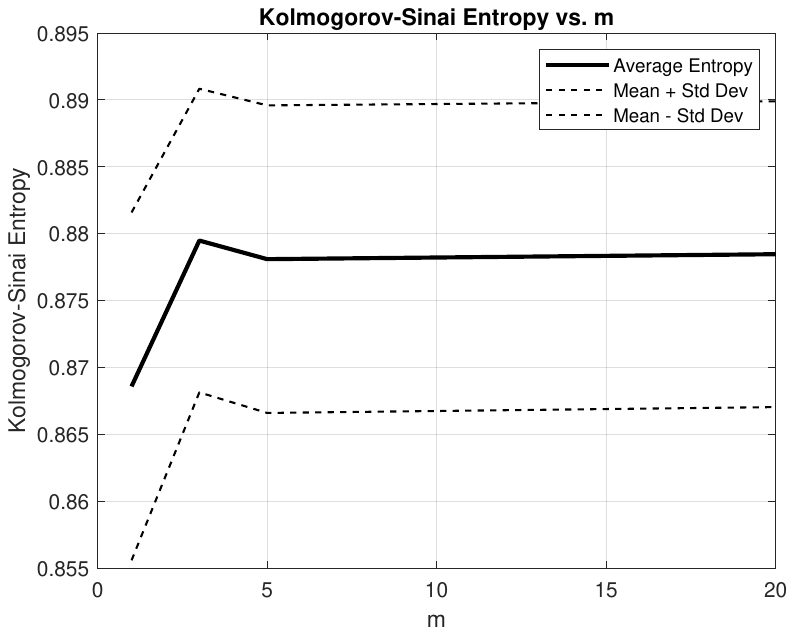}
		
	\end{figure}

	Figure \ref{fig:Hybrid} shows the percentage of adoption of the technology in the case of the absence of stubborn agents. 

	There is a positive dependence on the parameter of the preferential attachment. 
    %Also the trajectories are more regular. 

	As already stated above, K--S entropy is a measure of the unpredictability or chaos in a dynamical system. It quantifies the rate at which information about the system's state is lost over time, or, in other words, how chaotic or complex the system is.
	
	A low value of K--S entropy means that the system is more predictable, with less chaos or randomness, reflecting the strong predictability and regularity of the adoption dynamics.

    %In terms of social dynamics, this could indicate a system that reaches consensus or stability, where the behavior of agents (like opinions or adoption of innovations) becomes more uniform over time.
	
	A high value of K--S entropy indicates that the system is more chaotic or unpredictable, with a high degree of randomness, despite the knowledge of the past.

    %This could represent a system where the dynamics are highly volatile, such as in cases of polarization, continuous fluctuation in opinions, or where agents are consistently changing their behaviors in a disordered manner.
	
	%In the context of opinion dynamics or adoption models, a high entropy often corresponds to a state of disagreement or polarization, while a low entropy suggests a state of consensus or uniformity in opinions. Therefore,
    
%The plots on the left properly register how fast the dynamics approach the steady state.

The plot of the entropy shows the relevance of the network structure. When $m=1$, each new node connects through only one link, and the total number of links is minimal for the network model of BA. Therefore, a process starting at a random node has limited options to propagate compared to a BA network generated with preferential attachment parameter $m=3$, with a higher number of links, where more paths are available on the network. It is worth pointing out that after the peak in $m=3$ the entropy decreases. This means that although it is true that the number of links (and, consequently, the paths available for the propagation of the dynamics) increases as m increases, the diffusion becomes less erratic (i.e., the entropy decreases).  The reason can be seen observing the limit case of the complete network (obviously not BA), in which when everybody is connected to everybody else then the Bass dynamics end with very limited computational steps, since anyone in group $B$ is surely connected to some node in group $A$ because everybody is connected to everybody else. The value $m=3$ witnesses a specific “dangerous” moment for the dynamic to be maximally erratic.  Therefore, $m=3$ shows a peculiar value where the trajectories can be maximally random: at any time step, the increase of adopters of the new technology has the maximal variation, due to the network structure. Either a lower or a higher number of paths improves the smoothness of the process.
The node degree three has been shown to deserve attention also in the cascade correlation model of Elliot and Golub~\cite{elliott2014financial}. In their  Figure 2a,   a cascade failure of companies in a new market shows a peak of maximal fragility at the expected degree in the range [2.3, 4.3].

    %convergence to a common opinion, and add the information of the speed of convergence. 
    We see that already at $m=10$ the values are quite stable. This is in line with the proximity of the top curves in the left part of Figure \ref{fig:Hybrid}. 
	
	We remark that the absence of stubborn agents of $B$ type surely leads to the 100$\%$ adoption of the new technology. 
	Instead, each agent of type $A$ can be defined as stubborn, since they cannot float to $B$. The eventual presence of stubborn agents of $B$ type may lead to the maximal adoption of the technology at a fixed level, which can be lower than \mbox{100$\%$-percent} of $B$ type  nodes.
	 We add the remark that if we set a very limited number of stubborn agents of type $B$, it does not introduce relevant effects on the dynamics. On the other hand, in the limit case in which all the agents of type $B$ are stubborn the effect is equal to setting the parameters $p=0$ and $q=0$ (no conversion from B to A in any way), so the dynamics stops immediately at the first iteration. An intermediate number of stubborn not-adopter agents delays the adoption of new technology, and the effect of the delay depends on the positioning of the stubborn agents in the network, analogous to the studies reported in the literature on immunization strategies~\cite{pastor2002immunization,chen2008finding}.  Generally speaking, moving a hub from floater to stubborn has wider effects than having a stubborn agent  leave; moreover, both the network structure and the positioning of the stubborn agents influence the dynamics.

	\subsection{Second Case: Limit to the Maximal Size of the Groups}
\begin{figure}[]  
		%\centering
		\includegraphics[width=1.0\linewidth]{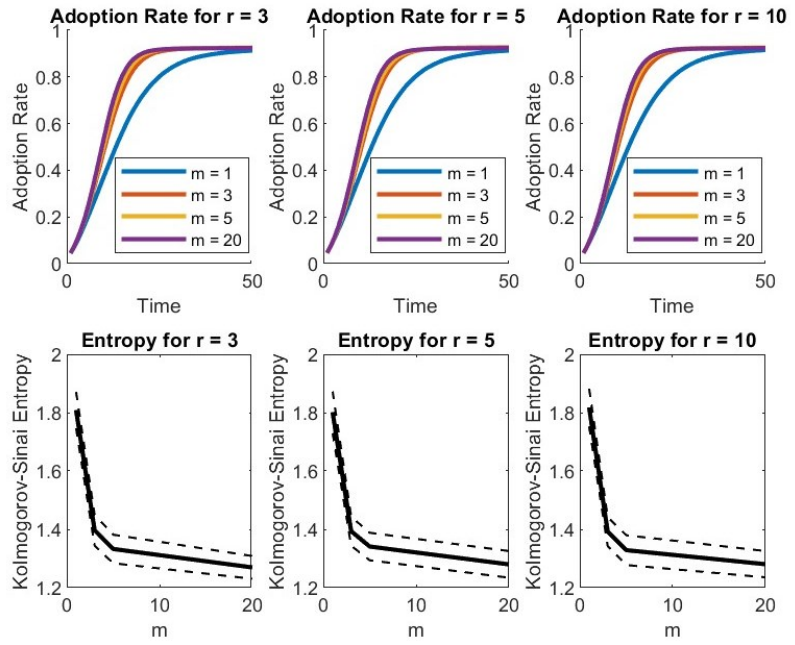}
		\caption{Plots with limit on the maximal sizes of the groups ($r=3, 5, 10$). 	}
		\label{fig:Unified}
	\end{figure}

	Still inspired by the work of Serge Galam, we tested the algorithm when the interaction for the majority rule is still driven by the network topology (only the first neighbours interact) but the maximal number considered is limited by $r$. Following some cases considered relevant by the literature, we selected $r=3, 5, 4$ as the maximal size of the groups. The constraints have no effect on the nodes poorly connected; the more the node is connected, the more the constraint on the group size is effective. In fact, out of the set of neighbours of each node, only $r$ are randomly extracted.
	The tests with maximal group size $r=3, 5, 10$ gave quite similar results (Figure \ref{fig:Unified}).
	
	The model analyzed differs from the Galam model in its absence of inflexible agents. Actually, since no switch back to the old technology is allowed, the presence of inflexibles on the old technology may eventually prevent 100$\%$ of the floaters from adopting the new  technology.
	
	\subsection{Possibility of Switching Back to the Old Technology}
	While the presence of inflexibles has been widely explored, we focus now on the presence of a charismatic competitor. 
	A charismatic competitor represents a focal agent who, through personal influence, attracts and persuades individuals. This effect is nonlinear, as the competitor's appeal can trigger network-wide opinion shifts. In the Galam framework, such competitors act similarly to biased influencers, promoting one opinion with a higher probability than random persuasion.
	We allow the possibility of switching back to the old technology only as a reaction to a charismatic competitor supporting the old technology~\cite{javarone2013network}.
	Charisma is relevant in several social contexts,
	since charismatic people usually exercise a strong influence over others.
	We assume that the node with the maximal degree (the hub) is the charismatic one. 
	It is assigned state $B$ and cannot change its state. If it is involved in the opinion formation process, it shifts back to  $B$ (i.e., the adoption of the old technology) the state of its neighbors with probability \( p \).
	
	The charismatic competitor introduces changes in the model; still, the effect remains different from the mean-field approach. 
	In fact, the charismatic competitor influences opinion shifts through personal appeal, and the mean-field effect assumes that individuals respond to the global average opinion:  each agent is influenced by the average global state of the system rather than localized peer interactions. This approach is widely used in statistical physics and opinion models to predict macroscopic behaviors.
	The presence of a charismatic competitor introduces a difference from the model presented \mbox{in~\cite{Galam2002,Galam2005,Galam2011,Galam2004,Galam1986,Galam2020,galam2007role}}.
	Serge Galam has already incorporated mean-field approaches in his studies of opinion dynamics. In~\cite{galam2010artifacts}, Galam (with Martins) investigates the dynamics of a one-dimensional Ising spin system using various local update rules, including the Galam majority rule. The study addresses discrepancies between analytical results and numerical simulations, discussing the implications of mean-field assumptions in modeling social behaviors.
	Furthermore, in~\cite{galam2022opinion} Galam introduces a probabilistic update formula within a Global Unifying Frame (GUF) to compare different opinion dynamics models. While the GUF differs from traditional mean-field approaches, it serves a similar purpose in simplifying complex interactions to predict macroscopic behaviors.
	
	By incorporating charismatic competitors in the hybrid Galam--Bass model, we can refine predictions about opinion shifts and innovation adoption. 
	
	When the charisma-driven persuasion rate is high, the effect of a charismatic competitor can approximate a mean-field dynamic by accelerating consensus formation.
	However, if the reach of the charismatic competitor is limited by the network topology, charisma-based influence remains quite distinct from mean-field homogenization.
	
			   \begin{table}\begin{center} % \fontsize{8pt}{8pt}\selectfont
\caption{Threshold of $p$ to reach the majority (50\%) of switch-back to the old technology due to the presence of a charismatic competitor.}\label{tab:charismatic} 

\begin{tabular}{|c||c|c|c|c|c|c|c|c|c|c|} \hline 
 $m$	&	$1$		&	$3$ 	&      $5$ 		&	$20$   \\ \hline   \hline
$p$	&	1.5\%	&0.5\%	&	0.2\%	&	0.1\%	\\
\hline	\end{tabular}
 \end{center}
 \end{table}

	%Todo: guardare il paper di galam del suo modello su BA e vedere se l'ha già fatto, altrimenti ripercorrere i passi verso la figura 7 del suo paper %https://arxiv.org/pdf/1901.09622

	We follow the approach suggested in~\cite{javarone2013network}, where each node reachable crossing two edges (i.e., the direct neighbors of the direct neighbors) is considered to be a neighbor.
    
     $M$ is the $0/1$ adjacency matrix of the network under examination, in order to implement the approach of~\cite{javarone2013network} the adjacency matrix to be used   becomes 
	\[
	M'=\text{sign}( M  + M^2)
	\]
	In fact, the matrix $M$ accounts for the direct neighbors, and the matrix $M^2$ has  no 	 null entries only if two nodes are connected by a path of length  2.  The sum $(M+M^2)$ has no null entries only when a node at line $i$ is either directly connected to a node column $j$ or the nodes $i$ and $j$ are connected through a path of length 2.   Since the network is not directed, the matrix is symmetric. Since we are not posing weights on the edges, the matrix needs to be a $0/1$ matrix; hence, the need to use the function \emph{ sign} 
$(\cdot)$.

	%disp(final_p_threshold);
	%Final p values for each m and network: 
	%         0         0         0         0         0         0         0         0         0         0
	%    0.0150    0.0450    0.0150    0.0050    0.0250    0.0150    0.0250         0.5         0.5    0.0150
	%    0.0040    0.0030    0.0100    0.0020    0.0030    0.0030    0.0020    0.0040    0.0020    0.0050
	%    0.0020    0.0050    0.0010    0.0040    0.0020    0.0060    0.0030    0.0020    0.0010    0.0030%quindi il minimo è: 
	%m   1    3    5 20
	%p  1.5\% 0.5\% 0.2\% 0.1\%

Interestingly, we can observe the behavior of the model in the presence of iterations with \( p \), i.e., the program runs iterations, increasing \( p \) by 0.001 each time, stopping when more than 50\% of the nodes have the same state as the competitor. Table \ref{tab:charismatic} shows the results. 
	The higher $m$ is, the more the connections to the hub, and the lower $p$ is to produce a shift back to the old technology.

	Future extensions may explore how different network structures mediate the transition from localized persuasion to mean-field behavior.

	\section{Conclusions}
	This paper elaborates on the opinion dynamics exploration by moving from a technological advancement-based perspective. We provide some paradigmatic examples of different patterns of innovation, also in light of the presence of key actors in the considered environment.
	
    In this regard, the hybrid Galam--Bass model provides a valuable framework for understanding the interplay between individual influence and collective dynamics in opinion formation and innovation diffusion. By integrating the effects of charismatic competitors instead that through mean-field approximations, the model captures the trade-off between localized persuasion mechanisms and global outcomes. This dual perspective enhances predictions of social and market behaviors, particularly in contexts such as political campaigns, viral marketing, and technological adoption. 
    
    Future research should explore how different network topologies, contrarian behaviors, and stubborn agents impact the transition from localized opinion shifts to the dominance of influencers, refining the balance between micro- and macro-level influences. In addition, a reasonable extension of the proposed setting is to create a setting merging the power of the Galam model with the prey--predator Lotka--Volterra dynamics (for a review, see~\cite{lv}), to discuss the application of sociophysics to innovator--imitator behaviors (on this topic,  refer to 
~\cite{cerquetiTFSC1,cerquetiTFSC2}). 

\bigskip
    {\bf Author contributions:} Conceptualization, G.R.; methodology, G.R. and R.C.; software, G.R.; validation, G.R., R.C., G.D., C.H., P.K., and M.A.; formal analysis, G.R., R.C., G.D., C.H., P.K., and M.A.; investigation, G.R., R.C., G.D., C.H., P.K., and M.A.; writing---original draft preparation, G.R., R.C., G.D., C.H., P.K., and M.A.; writing---review and editing, G.R., R.C., G.D., C.H., P.K., and M.A.; supervision, G.D., C.H., P.K., and M.A. 
    All authors have read and agreed to the published version of the~manuscript.

{\bf  Funding:} This research received partial funding under the project ‘’A better understanding of socioeconomic
systems using Quantitative Methods from Physics”, funded by the European Union—Next
generation EU and the Romanian Government under the National Recovery and Resilience Plan for
Romania, contract no.760034/23.05.2023, code PNRR-C9-I8-CF 255/29.11.2022, through the Romanian
Ministry of Research, Innovation and Digitalization, within Component 9, ‘’Investment I8” \& the
European Union–Next Generation EU, Project PRIN2022 “Networks: decomposition, clustering
and community detection” code: 2022NAZ0365 – CUP H53D2300 2510006 \& Departmental project
Sapienza University of Rome "Leaving No One Behind: Methods for Sustainability" Protocollo nr.:
 RD124190DA1146AA – CUP: B83C24007080005.
	
{\bf Data availability: } All the data have been simulated. The necessary information for
generating them through computer software is contained within the article. 

	{\bf  Acknowledgments:} GR acknowledges the scientific support of INdAM/GNFM.

	{\bf Conflicts of interest}  The authors declare no conflicts of interest.

%\begin{adjustwidth}{-\extralength}{0cm}
%\centering %% If there is a figure in wide page, please release command \centering, for Table, ``\textwidth" should be ``\fulllength"

%\reftitle{References}	

 %%%%%%%%%%%%%%%%%%%%%%%%%%%%%%%%%%%%%%%%%%
%\end{adjustwidth}


\begin{thebibliography}{999}

	%\bibliographystyle{plain}  % Stile di citazione (puoi cambiarlo in base alle tue esigenze)
	%\bibliography{entropy-3740600} 


\bibitem%[Galam(2002)] 1.
{Galam2002}
 Galam, S.  
\newblock Minority opinion spreading in random geometry.
\newblock {\em  Eur. Phys. J. B-Condens. Matter Complex Syst.} {\bf 2002}, {\em 25},~403--406.

\bibitem%[Galam(2011)]2.
{Galam2011}
Galam, S.
\newblock What is sociophysics about? In {\em Sociophysics: A Physicist's
  Modeling of Psycho-Political Phenomena}; Springer:   Berlin/Heidelberg, Germany,
  2011; pp. 3--19.

\bibitem%[Galam(1986)]3.
{Galam1986}
Galam, S.
\newblock Majority rule, hierarchical structures, and democratic
  totalitarianism: A statistical approach.
\newblock {\em J. Math. Psychol.} {\bf 1986}, {\em
  30},~426--434.

\bibitem%[Galam(2004)]4.
{Galam2004}
Galam, S.
\newblock The dynamics of minority opinions in democratic debate.
\newblock {\em Phys. A: Stat. Mech. Its Appl.} {\bf
  2004}, {\em 336},~56--62.

\bibitem%[Galam(2005)]5.
{Galam2005}
Galam, S.
\newblock Heterogeneous beliefs, segregation, and extremism in the making of
  public opinions.
\newblock {\em Phys. Rev. E—Stat. Nonlinear Soft Matter Phys.} {\bf 2005}, {\em 71},~046123.


\bibitem%[Bass(2004)] 6.
{Bass2004}
Bass, F.M.
\newblock Comments on ``A New Product Growth for Model Consumer Durables'' by
  Frank M. Bass.
\newblock {\em Manag. Sci.} {\bf 2004}, {\em 50},~1833-- 1836. %\newblock {\url{https://doi.org/10.1287/mnsc.1040.0162}}.

\bibitem%[Mahajan and Muller(1990)] 7.
{Mahajan1990}
Mahajan, V.; Muller, E.
\newblock New Product Diffusion Models in Marketing: A Review and Directions
  for Research.
\newblock {\em J. Mark.} {\bf 1990}, {\em 54},~1--26.
%\newblock {\url{https://doi.org/10.1177/002224299005400201}}.

\bibitem%[Tigert and Farivar(1981}8.
{tigert}
Tigert, D.; Farivar, B.
\newblock The Bass new product growth model: A sensitivity analysis for a high
  technology product.
\newblock {\em J. Mark.} {\bf 1981}, {\em 45},~81--90.

\bibitem{VRAC2015} % 9.
Varela, L. M.;  Rotundo, G.;  Ausloos, M.; Carrete, J. Complex network analysis in socioeconomic models. In {\it Complexity and Geographical Economics: Topics and Tools}; Springer: Berlin/Heidelberg, Germany, 2015; pp. 209–245.
% https://doi.org/10.1007/978-3-319-12805-4_9

\bibitem%[Barabási and Albert(1999)] 10.
{barabasi1999-1}
Barabási, A.L.; Albert, R.
\newblock Emergence of Scaling in Random Networks.
\newblock {\em Science} {\bf 1999}, {\em 286},~509--512.

\bibitem%[Pastor-Satorras et~al.(2015){Pastor-Satorras, Castellano, Van~Mieghem, and Vespignani] 11.
{Vespignani2006-1}
Pastor-Satorras, R.; Castellano, C.; Van~Mieghem, P.; Vespignani, A.
\newblock Epidemic processes in complex networks.
\newblock {\em Rev. Mod. Phys.} {\bf 2015}, {\em 87},~925--979.

\bibitem%[Barab{\'a}si(2002)] 12.
{Barabasi2002}
Barab{\'a}si, A.L.
\newblock {\em The New Science of Networks}; Perseus Books Group: Cambridge,
  MA,  USA,  2002.

\bibitem%[Kolmogorov(1958)] 13.
{kolmogorov1958invariant}
Kolmogorov, A.N.
\newblock A new metric invariant of transient dynamical systems and
  automorphisms in Lebesgue spaces.
\newblock {\em Dokl. Akad. Nauk SSSR} {\bf 1958}, {\em 119},~861--864.

\bibitem%[Sinai(1959)] 14.
{sinai1959entropy}
Sinai, Y.G.
\newblock On the concept of entropy for a dynamic system.
\newblock {\em Dokl. Akad. Nauk SSSR} {\bf 1959}, {\em 124},~768--771.

\bibitem%[Mackay et~al.(2015)Mackay, Hailu, Ramirez-Elizondo, and Bauer] 15.
{giersch2016usbc-1}
Mackay, L.; Hailu, T.; Ramirez-Elizondo, L.; Bauer, P.
\newblock Towards a DC distribution system-opportunities and challenges.
\newblock In Proceedings of the 2015 IEEE First International Conference on DC
  Microgrids (ICDCM),  Atlanta, GA, USA, 7--10 June 2015;
 IEEE:  Piscataway, NJ, USA, 2015, pp. 215--220.

\bibitem%[Mackay et~al.(2017)Mackay, Blij, Ramirez-Elizondo, and  Bauer] 16.
{benedict2009usb-1}
Mackay, L.; Blij, N.H.v.d.; Ramirez-Elizondo, L.; Bauer, P.
\newblock Toward the universal DC distribution system.
\newblock {\em Electr. Power Compon. Syst.} {\bf 2017}, {\em
  45},~1032--1042.

\bibitem%[Cheng et~al.(2019)Cheng, Yu, Jiang, Shi, Tan, and  Zhang] 17.
{buchanan2018role-1}
Cheng, L.; Yu, T.; Jiang, H.; Shi, S.; Tan, Z.; Zhang, Z.
\newblock Energy internet access equipment integrating cyber-physical systems:
  Concepts, key technologies, system development, and application prospects.
\newblock {\em IEEE Access} {\bf 2019}, {\em 7},~23127--23148.

\bibitem%[Vellequette(2023)] 18.
{sobel2017usbc-1}
Vellequette, L.P.
\newblock Plugging into the future; Some automakers slow to switch to faster  USB-C ports.
\newblock {\em Automot. News}  {\bf 2023}, {\em 98}.

\bibitem%[Jian et~al.(2012)Jian, Syed-Abdul, Sood, Lee, Hsu, Ho, Li, and Wen] 19.
{berkman2019usb-1}
Jian, W.S.; Syed-Abdul, S.; Sood, S.P.; Lee, P.; Hsu, M.H.; Ho, C.H.; Li, Y.C.;  Wen, H.C.
\newblock Factors influencing consumer adoption of USB-based Personal Health  Records in Taiwan.
\newblock {\em BMC Health Serv. Res.} {\bf 2012}, {\em 12},~ 277.

\bibitem%[Cronin et~al.(2022)Cronin, Gao, Wang, and Cotton] 20.
{barrett2020usb-1}
Cronin, P.; Gao, X.; Wang, H.; Cotton, C.
\newblock Time-print: Authenticating USB flash drives with novel timing  fingerprints.
\newblock In Proceedings of the 2022 IEEE Symposium on Security and Privacy
  (SP),  San Francisco, CA, USA, 23--25 May 2022; 
 IEEE:  Piscataway, NJ, USA,  2022, pp. 1002--1017.

\bibitem%[Mom(2013)] 21.
{sivak2014early-1}
Mom, G.
\newblock {\em The Electric Vehicle: Technology and Expectations in the
  Automobile Age}; JHU Press:  Baltimore, MD, USA,  2013.

\bibitem%[Yakovlev(2022)] 22.
{hirsch1999rise-1}
Yakovlev, V.F.
\newblock Early electric vehicle charging: A survey.
\newblock {\em Int. J. Electr. Hybrid Veh.} {\bf
  2022}, {\em 14},~219--230.

\bibitem%[Gordon(2024)] 23.
{buchanan2009electric-1}
Gordon, C.E.
\newblock Putting the Car Before the Horse: The Diffusion of the Automobile and
  the Rise of Technocratic Primacy.
\newblock {\em Histories} {\bf 2024}, {\em 4},~487--507.

\bibitem%[Tollefson(2017)] 24.
{morganlewis2025-1}
Tollefson, J.
\newblock Trump pulls United States out of Paris climate agreement.
\newblock {\em Nature} {\bf 2017}, {\em 546}.

\bibitem%[Fiorino(2025)] 25.
{guardian2025-1}
Fiorino, D.J.
\newblock What Does a Second Trump Term Mean for US Environmental Policy?
\newblock {\em Policy Q.} {\bf 2025}, {\em 21},~46--49.

\bibitem%[Bass(1969)] 26.
{Bass1969}
Bass, F.M.
\newblock A New Product Growth for Model Consumer Durables.
\newblock {\em Manag. Sci.} {\bf 1969}, {\em 15},~215--227.
%\newblock {\url{https://doi.org/10.1287/mnsc.15.5.215}}.

\bibitem%[Galam and Cheon(2020)]
{Galam2020}
Galam, S.; Cheon, T.
\newblock Tipping Points in Opinion Dynamics: A Universal Formula in Five
  Dimensions.
\newblock {\em Front. Phys.} {\bf 2020}, {\em 8},~566580.
%\newblock {\url{https://doi.org/10.3389/fphy.2020.566580}}.

\bibitem%[Galam(2023)]
{galam2023unanimity}
Galam, S.
\newblock Unanimity, coexistence, and rigidity: Three sides of polarization.
\newblock {\em Entropy} {\bf 2023}, {\em 25},~622.

\bibitem%[Jacobs and Galam(2019)]
{jacobs2019two}
Jacobs, F.; Galam, S.
\newblock Two-opinions-dynamics generated by inflexibles and non-contrarian and
  contrarian floaters.
\newblock {\em Adv. Complex Syst.} {\bf 2019}, {\em 22},~1950008.

\bibitem%[Li and Zehmakan(2023)]
{li2023graph}
Li, S.; Zehmakan, A.N.
\newblock Graph-based generalization of Galam model: Convergence time and
  influential nodes.
\newblock {\em Physics} {\bf 2023}, {\em 5},~1094--1108.

\bibitem%[Pastor-Satorras and Vespignani(2002)]
{pastor2002immunization}
Pastor-Satorras, R.; Vespignani, A.
\newblock Immunization of complex networks.
\newblock {\em Phys. Rev. E} {\bf 2002}, {\em 65},~036104.

\bibitem%[Bertotti and Modanese(2019)]
{bertotti2019bass}
Bertotti, M.L.; Modanese, G.
\newblock The Bass Diffusion Model on Finite Barabasi-Albert Networks.
\newblock {\em Complexity} {\bf 2019}, {\em 2019},~6352657.

\bibitem%[Jiang et~al.(2006)Jiang, Bass, and Bass]
{jiang2006virtual}
Jiang, Z.; Bass, F.M.; Bass, P.I.
\newblock Virtual Bass Model and the left-hand data-truncation bias in
  diffusion of innovation studies.
\newblock {\em Int. J. Res. Mark.} {\bf 2006}, {\em
  23},~93--106.

\bibitem%[Javarone(2014)]
{javarone2013network}
Javarone, M.A.
\newblock Network strategies in election campaigns.
\newblock {\em J. Stat. Mech. Theory Exp.} {\bf
  2014}, {\em 2014},~P08013.

\bibitem%[Elliott et~al.(2014)Elliott, Golub, and Jackson]
{elliott2014financial}
Elliott, M.; Golub, B.; Jackson, M.O.
\newblock Financial networks and contagion.
\newblock {\em Am. Econ. Rev.} {\bf 2014}, {\em 104},~3115--3153.

\bibitem%[Chen et~al.(2008)Chen, Paul, Havlin, Liljeros, and  Stanley]
{chen2008finding}
Chen, Y.; Paul, G.; Havlin, S.; Liljeros, F.; Stanley, H.E.
\newblock Finding a better immunization strategy.
\newblock {\em Phys. Rev. Lett.} {\bf 2008}, {\em 101},~058701.

\bibitem%[Galam and Jacobs(2007)]
{galam2007role}
Galam, S.; Jacobs, F.
\newblock The role of inflexible minorities in the breaking of democratic
  opinion dynamics.
\newblock {\em Phys. A: Stat. Mech. Its Appl.} {\bf  2007}, {\em 381},~366--376.

\bibitem%[Galam and Martins(2010)]
{galam2010artifacts}
Galam, S.; Martins, A.C.R.
\newblock Artifacts of opinion dynamics at one dimension.
\newblock {\em arXiv} {\bf 2010},  arXiv:1012.2283.

\bibitem%[Galam(2022)]
{galam2022opinion}
Galam, S.
\newblock Opinion Dynamics and Unifying Principles: A Global Unifying Frame.
\newblock {\em Entropy} {\bf 2022}, {\em 24},~1201.

\bibitem%[Wangersky(1978)]
{lv}
Wangersky, P.J.
\newblock Lotka-Volterra population models.
\newblock {\em Annu. Rev. Ecol. Syst.} {\bf 1978}, {\em  9},~189--218.

\bibitem%[Cerqueti et~al.(2015)Cerqueti, Tramontana, and Ventura]
{cerquetiTFSC1}
Cerqueti, R.; Tramontana, F.; Ventura, M.
\newblock On the coexistence of innovators and imitators.
\newblock {\em Technol. Forecast. Soc. Chang.} {\bf 2015}, {\em  90},~487--496.

\bibitem%[Cerqueti et~al.(2016)Cerqueti, Quaranta, and Ventura]
{cerquetiTFSC2}
Cerqueti, R.; Quaranta, A.G.; Ventura, M.
\newblock Innovation, imitation and policy inaction.
\newblock {\em Technol. Forecast. Soc. Chang.} {\bf 2016}, {\em  111},~22--30.

\end{thebibliography}
\end{document}